# Ancient River Morphological Features on Mars versus Arizona's Moenkopi Plateau

AN EARTH-ANALOG INVESTIGATION

ANTONIO J. PARIS & LAURENCE A. TOGNETTI

PLANETARY SCIENCES, INC.

**ABSTRACT**

Mars is currently at the center of scientific debate regarding proposed ancient river morphological landscapes on the planet. An increased curiosity in the geomorphology of Mars and its water history, therefore, has led to an effort to better understand how those landscapes formed. Many studies, however, consist of patchwork investigations that have not thoroughly examined proposed ancient fluvial processes on Mars from an Earth-analog perspective. The purpose of this investigation, therefore, is to compare known fluvial features on Moenkopi Plateau with proposed paleopotamologic features on Mars. The search for analogs along the Moenkopi Plateau was due to the similarities in fluvial erosion, influenced and modified by eolian (wind) activity, primarily from Permian through Jurassic age. By analyzing orbital imagery from two cameras onboard NASA's Mars Reconnaissance Orbiter (MRO) - the High-Resolution Imaging Science Experiment (HiRISE) and the Context Camera (CTX) - and paralleling it with imagery obtained from the U.S. Geological Survey and an unmanned aircraft operating over the Moenkopi Plateau, this investigation identified similar fluvial morphology. We interpret, therefore, that the same fluvial processes occurred on both planets, thereby reinforcing the history of water on Mars.

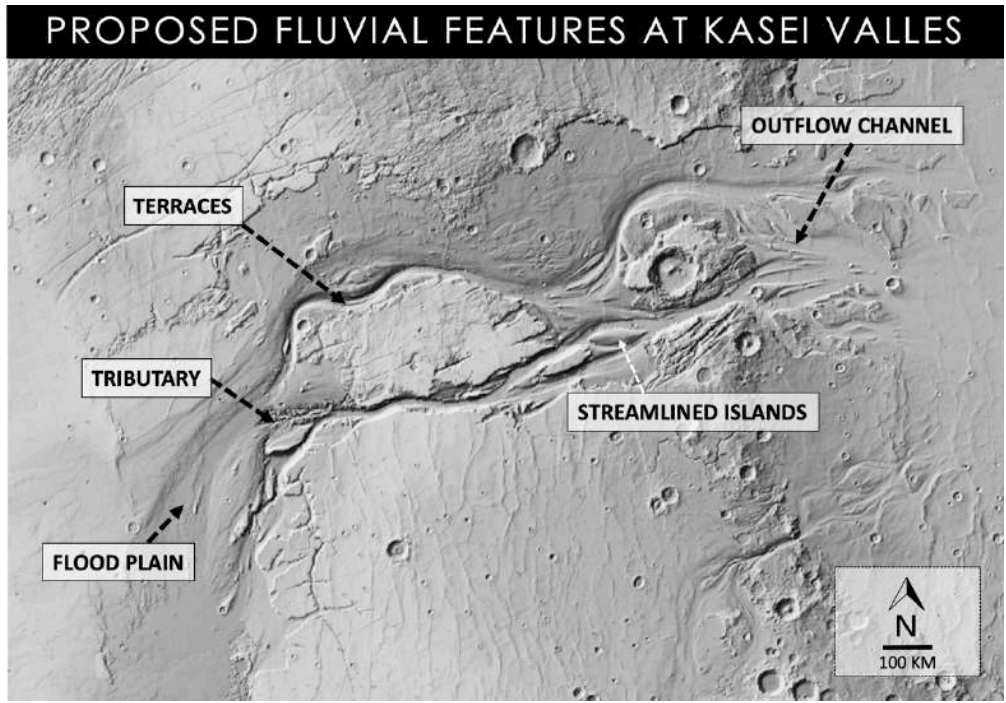

Figure 1: Proposed fluvial features created by the flow of water. Source: NASA Mars Orbiter Laser Altimeter

## PROPOSED FLUVIAL FORMATIONS ON MARS

The history of water on Mars is a matter of contention, and one of the essential questions planetary scientists are attempting to unravel. The *Kasei Valles*, for illustration, is a vast system of canyons in the *Mare Acidalium* and *Lunae Palus* quadrangles on Mars, centered at latitude 24.6° N and longitude 65.0° W (Figure 1). The canyons are ~1,580 km long and represent one of the largest proposed outflow channel systems on the planet.[1] Numerous studies have attempted to interpret the troughs and valleys of *Kasei Valles* as incontrovertible outflow channels, but their history and origin remain ambiguous. Surface features in the region appear to represent an outflow area that could have been the result of catastrophic flooding millions of years ago. Proposed paleopotamologic features in *Kasei Valles,* such as tributaries, terraces, flood plains, and streamlined islands, appear analogous to known fluvial features found on the Moenkopi Plateau, which were shaped by the flow of water. Others proposed that glacial action or volcanic activity produced the paleopotamologic features on Mars, rather than the flow of liquid water.[2]

## MOENKOPI PLATEAU

This investigation focuses on the Moenkopi Plateau in northeastern Arizona (Figure 2). The area extends from the Little Colorado River northeastward to the summit, covering 484 km². Elevations range from 1,280 m at the Little Colorado River to 1,700 m on the plateau. The Adeii Eechii Cliffs, an erosional scarp, demarcates the southwest edge of the Moenkopi Plateau. Other major erosional scarps to the southwest include the Red Rock Cliffs and Ward Terrace.[3]

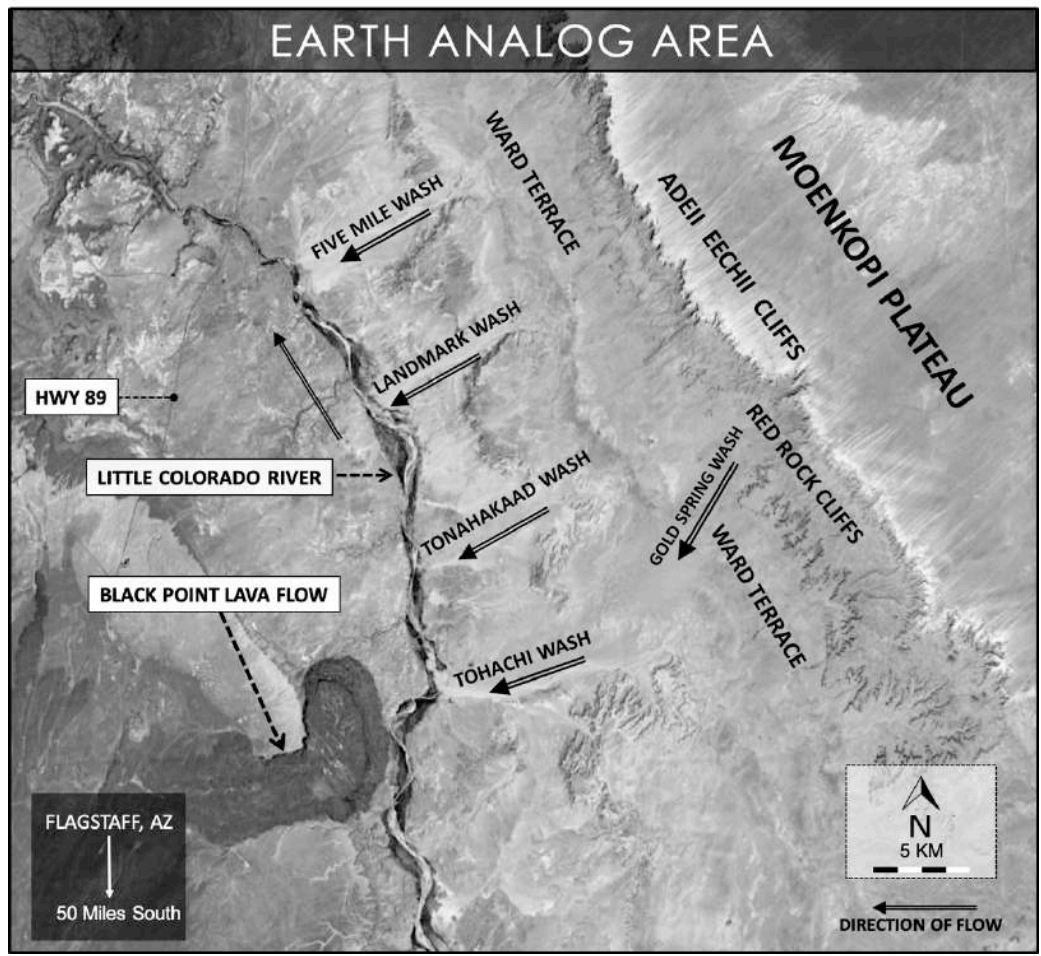

Figure 2: The Moenkopi Plateau. Source: USGS

Geomorphic interactions between eolian and fluvial processes since the late Pleistocene are reflected by drainage patterns on northeasterly plunging sedimentary rocks and by the northeastward withdrawal of cuestas along the southwest boundary of the plateau.[3] Tributary drainages, such as the Five Mile, Landmark, Tonahakaaad, Tohachi, and Gold Spring Wash, flow southwestward from the edge of the plateau toward the Little Colorado River. Subsequent fluvial drainages, moreover, are entrenched within resistant strata near the base of retreating scarps, leaving a visual record of scarp retreat and development during the past ~2.4 million years.[4] During the Pliocene, wind-swept sand from indigenous sedimentary strata was transported from Little Colorado in the southwest onto the Moenkopi Plateau in the northeast, opposite of the direction of fluvial flow.

As noted above, the age of present fluvial and eolian deposits in the analog area is Holocene and Pleistocene, undivided.[5] Geologically, the plateau consists of exposed sedimentary rocks of Permian through Jurassic; volcanic basalt deposits from the Black Point lava flow; and surficial deposits consisting of sand dunes, sand sheets and landside deposits (Figure 3). Sedimentary rocks that consist of silica-cemented sandstone, interbedded limestone, and multi-colored shale plunge to the northeast and form northwest-trending ledges and cliffs. The bedrock in the area of study, moreover, has been eroded by streams, winds, and a copious supply of loose sediment available for redeposition.[6]

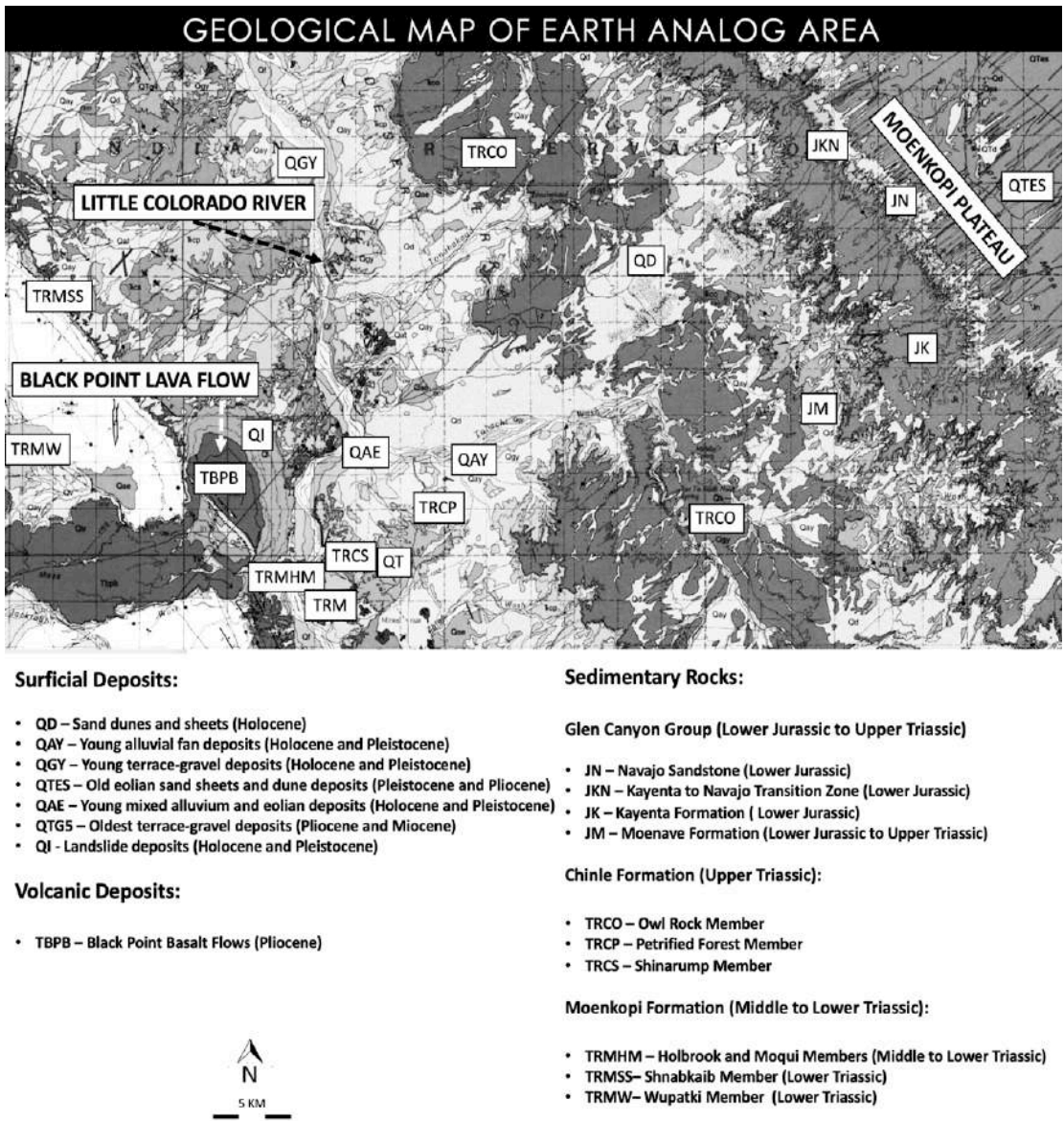

Figure 3: Geological Survey of the Moenkopi Plateau. Source: USGS and Planetary Sciences, Inc.

# GEOMORPHOLOGY OF FLUVIAL SYSTEMS ON EARTH

Fluvial systems, the most significant geomorphic agent on Earth, are primarily dominated by streams and rivers. For millions of years, fluvial processes have sculpted, eroded, and transported sediment to create new landforms. The watershed or drainage basin is a fundamental landscape component in fluvial geomorphology and consists of a parent river and its tributaries.[7] The rivers, streams, and the depositional and erosional behavior of fluvial systems produce a variety of geomorphic topographies along the floodplain, such as meandering systems, tributaries, terraces, oxbows, confluence, braiding, cut banks, point bars, streamlined islands, and terraces (Figure 4). Erosion, which originates from the power and consistency of the current, can affect the formation of the river's path.[8] Furthermore, the availability, rate of deposition, size, and composition of sediment moving through the channel will shape and change the direction of the river over time.[9] Dunes and sand sheet deposits due to eolian activity further contribute to the morphology of the Moenkopi Plateau.

Today, stream deposits between the Little Colorado River and the Moenkopi Plateau consist mostly of mud, silt sand, and gravel. Throughout seasonal dry spells in late spring and early summer, sand and silt travel in the wind from stream channels, primarily Tohachi Wash (Figure 2), onto adjacent flood plains.[3] Most of the sand occurs in sheets that advance northeasterly. Where the sand is relatively thin or sparse, it forms well-defined barchan or parabolic dunes while thicker sand sheets form complex dunes.[10] Some of the sand blown by the wind from washes, however, is transported by streams back into the washes. This recycling takes place in all drainages east of the Little Colorado River regardless of their orientation.[11]

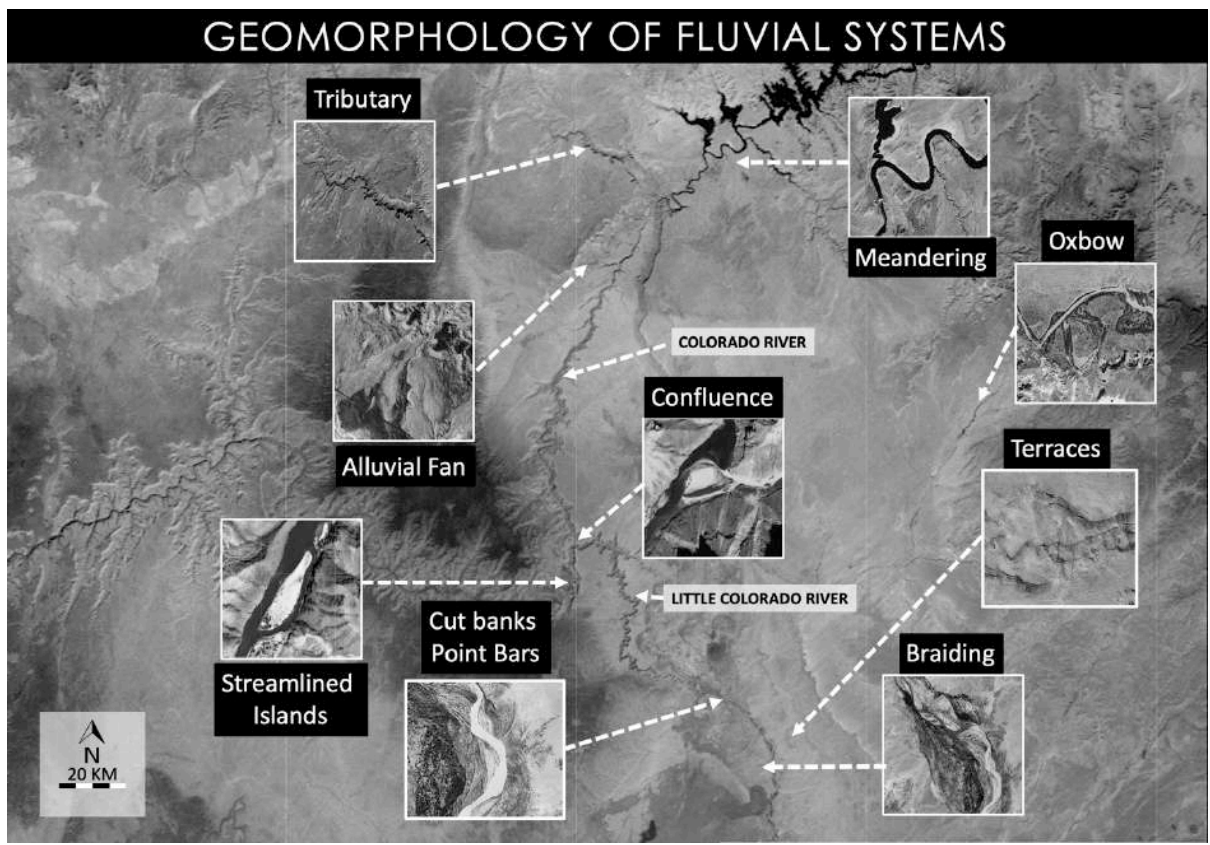

Figure 4: Examples of Fluvial Morphology. Source: USGS and Planetary Sciences, Inc.

## DATA COLLECTION:

## ORBITAL IMAGERY OF MARS AND IMAGERY OF THE MOENKOPI PLATEAU

The NASA HiRISE and CTX imagery used in this investigation were available through NASA's Planetary Data System (PDS) and the Lunar and Planetary Laboratory, University of Arizona. HiRISE can see the surface of Mars with a high-resolution capability up to ~30 centimeters per pixel, while the MRO CTX camera can observe at ~6 meters per pixel.[12] MRO also observes the Martian surface earlier in the day; thus, more geomorphic features are evident with partially sunlit floors.[13] A partially sunlit floor allows planetary scientists to identify specific features characteristic to paleofluvial action and floor morphologies, such as tributaries, terraces, flood plains, and streamlined islands. The data for Martian surface composition and properties came from the Thermal Emission Spectrometer (TES) onboard the Mars Global Surveyor spacecraft, which is accessible through the Java Mission-planning and Analysis for Remote Sensing (JMARS). The database is a geospatial information system (GIS) developed by ASU's Mars Space Flight Facility to provide mission planning and data-analysis tools to NASA scientists and instrument team members. Global mineral abundance maps were derived using atmospherically corrected TES spectral data and a suite of 36 endmembers, and interpolation used between adjacent orbit tracks.[14] The MRO images (HiRISE and CTX) were then compared with imagery obtain through the use of a crewless aerial vehicle (UAV) operated by Planetary Science, Inc. *in situ* on the Moenkopi Plateau. The UAV offered a powerful camera on a 3-axis stabilized gimbal that recorded video at 4k resolution up to 60 frames per second and featured real-glass optics that captured aerial imagery at 12 megapixels from an altitude up to 800 m and a range of 7 km.[15] Data obtained from the U.S. Geological Survey Earth Explorer (EE) imagery interface for high altitude aerial imagery also provided remote sensing inventory of the Moenkopi Plateau.

## IMAGERY ANALYSIS & INTERPRETATION

In Appendix 1, we present a series of orbital imagery of *proposed* paleopotamologic formations on Mars alongside aerial imagery of *known* fluvial morphology on the Moenkopi Plateau. The images infer that known fluvial processes that occur on the Moenkopi Plateau also took place on Mars early in its history. As the two planets are comparable compositionally, their rocks and minerals have similar names.[16] The shape and size of geomorphic processes on Mars, however, depend mostly on a set of environmental conditions and properties dissimilar to Earth. Lesser atmospheric pressure altered the scattering of material, and lower gravity facilitated wider dispersion.[17] Consequently, fluvial artifacts on Mars are larger than their terrestrial analogs.

## ANALOG 1: TRIBUTARY

A tributary is a stream or river that does not flow directly into a sea or ocean. In this fluvial setting, as exhibited in Figure 5, the water in both tributaries flowed into a more significant stream, parent river, or channel. These waterways, including the main stem river, drain the neighboring drainage basin of its surface water and groundwater, leading the water out into an ocean.[18] The tributary located in the western region of *Nilus Mensae* (latitude 22.102° N and longitude 287.570° E) is of the ~4 billion-year Noachian age, undivided, and characterized by canyons and channel floors similar to the tributaries in the Moenkopi Plateau. The *Nilus Mensae* area is a highland transition unit dominated by complex admixtures of impact, sedimentary, and volcanic rocks; it primarily consists of basalt, andesite, and traces of feldspar, hematite, and quartz.[19] The tributary investigated in the Moenkopi Plateau (latitude 35° 43' 45"

N and longitude 111° 18' 51" W) is of Holocene to Pleistocene age. The area is dominated by deposits of basaltic ash fragments, quartz, feldspar, gray-brown chert, quartz sandstone grains, and quartz with iron inclusions.[3]

## ANALOG 2: CONFLUENCE

A confluence or conflux occurs where two or more flowing bodies of water intersect to form a single river or where a tributary or wash joins a parent river or channel.[20] The natural flow of confluences gives rise to hydrodynamic patterns, such as mixing layers, stagnation zones, and shear layers, which, in some instances, can be identified from orbital and aerial imagery. Denser materials transported by either flow, such as large pebbles and rocks, sink to the main river bottom at or downstream of the confluence and amalgamate into deposits.

*Dao Vallis*, a proposed outflow channel on Mars, is of the warmer wetter ~3 billion year Hesperian age and runs southwestward into *Hellas Planitia* from the southern slopes of the volcano *Hadriacus Mons*. It and its tributary, *Niger Vallis*, extend for about 1,200 km.[21] Geologically, the area along the channel is comprised of planar deposits meters to tens of meters thick and tens to hundreds of kilometers across; flood lavas sourced from a regional fissure; and vent systems and lobate scarps are also typical.[22] Some proposed that *Dao Vallis* received water when hot magma from *Hadriacus Mons* melted vast amounts of ice in the frozen ground, released in massive outburst floods.[23] In one particular paleofluvial setting at *Dao Vallis* (latitude -36.804° S and longitude 89.990° E), settled mixing layers, stagnation zone, and shear layering can be identified (Figure 6). Telemetry from TES indicates an abundance of basalt, feldspar, and traces of quartz and hematite.[19] Conversely, the confluence examined on the Moenkopi Plateau (latitude 35° 43' 10" N and longitude 111° 08' 06" W) likewise arose from the convergence of two streams flowing southwestward from the Adeii Eechii Cliffs. The seasonal wash consists of a young mixture of alluvium fan deposits (Holocene and Pleistocene) dominated by clear quartz, milky quartz, light-red quartz, blue-gray chert, and fragments of basaltic ash fragments and schist.[3] Owl Rock, Chinle Formation (Upper Triassic), borders the rivers.

## ANALOG 3: STREAMLINED ISLANDS

Streamlined or teardrop-shaped islands stand in the beds of most large outflow channels on Mars. These islands mark where rock outcrops made obstructions that successfully resisted the floods.[24] Along the flanks of some streamlined islands, ledges or benches develop. These indicate particularly resistant strata or where the flood maintained a depth long enough to erode the ledge or bench. After floodwaters divide around the obstruction, they progressively erode the ground behind it.[25]

Lying east of the volcanic region of *Tharsis*, *Kasei Valles* is the most significant proposed outflow channel on Mars. Similar to fluvial systems on the Moenkopi Plateau, the channels of *Kasei Valles* appear to have been carved by liquid water, possibly during massive floods that originated in tectonic and volcanic activity in *Tharsis*.[24] Though larger than their terrestrial counterpart, streamlined islands are abundant at *Kasei Valles*. A particular streamlined island on *Kasei Valles* (latitude 24.736° N and longitude 311.401° E) unmistakably exhibits morphology consistent with the floodwater divide, including ledges and benches (Figure 7). This streamlined island of Noachian age is dominated by basalt, andesite, and traces of sulfate, feldspar, and carbonate. Additionally, the island is gradational with *Chryse Planitia* - a posited large lake or an ocean during the *Hesperian* or the subsequent cold dry *Amazonian* period.[19] The analogous streamlined island at the Moenkopi Plateau (latitude 35° 46' 08" N and longitude 111° 19' 30" W) is a stream-channel of Holocene age (Figure 7). The composition of bedrock samples along the streamlined island mainly consists of basaltic ash fragments, milky quartz, clear quartz, white quartz sandstone grains, and traces of magnetite and siltstone mud curls.[3] This analog, moreover, also exhibits ledges and benches consistent with fluvial erosion.

## ANALOG 4: TERRACES

Terraces commonly appear on the banks of channels and rivers. Their presence indicates sufficient fluvial activity for erosion before the water receded. Terraces also could have been shaped by several layers of strata that resisted erosion better than the layers above or below.[24] There are a variety of terraces of different sizes along the channels of *Kasei Valles*. The channel located at latitude 23.829° N and longitude 295.544° E distinctly exhibits terracing steps consistent with fluvial morphology (Figure 8). These Noachian-age terraces feature basalt, andesite and traces of pyroxene, quartz, sulfate, olivine, sheet silicates, feldspar, carbonates, and hematite.[19] The terraces investigated on the Moenkopi Plateau (latitude 35° 45' 20" N and longitude 110° 56' 18" W) likewise exhibit several layers of erosion due to numerous stages of water flow. These terraces are part of the Navajo Sandstone, Glen Canyon Group laid down in the Lower Triassic and possibly Early Triassic.[6] The wide range of colors exposed along the Navajo Sandstone reflects a long history of modification by groundwater and other subsurface fluids over the last 190 million years.[26] The diverse colors originate from the presence of variable mixtures and amounts of hematite, goethite, and limonite that fill the pore space of the quartz sand in the Navajo Sandstone.[27]

## ANALOG 5: ALLUVIAL FANS

Alluvial fans are triangular-shaped deposits of material transported by water. They are an example of unconsolidated sediment and tend to form in elevated regions with a rapid change in slope from a high to a low gradient.[28] The water flow transporting the sediment runs at a relatively high velocity due to the steep slope, which is why coarse material can remain in the stream. When the gradient decreases into a flat area, the fluvial action loses the energy it needs to transport the sediment further. The deposit eventually spreads out, forming an alluvial fan.[29]

*Capri Chasma* lies in the eastern portion of the *Valles Marineris*, the largest known canyon system in the Solar System. Deeply incised canyons such as *Capri Chasma* are exceptional targets for examining alluvial morphology, as many of the walls reveal distinct types of bedrock. *Capri Chasma is a* late Noachian highland comprised of undifferentiated impact, volcanic, fluvial, and basin material.[6] The alluvial fan located on *Capri Chasma* at latitude -13.354° S and longitude 308.285° E reveals how material was transported by water from a minor tributary to form the fan (Figure 9). Telemetry from TES at this particular fan shows abundant basalt and andesite with traces of quartz, sulfate, olivine, sheet silicates, feldspar, carbonates, and hematite.[19] The alluvial fan examined east of the Moenkopi Plateau (latitude 36° 12' 41" N and longitude 11123' 28" W) exhibits fluvial morphology analogous with the alluvial fan on *Capri Chasma*. In this geologically setting, water flowed from the top of the plateau through a series of stratigraphy (Kayenta Formation, Wingate Sandstone, and Chinle Formation) exposing fluvial siltstone, fine-grained silty sandstone with interbedded purplish-red shale and authigenic quartz.[3]

## ANALOG 6: BRAIDING

A braided river is a network of river channels separated by small, often temporary, islands called braid bars. Braiding tends to occur in steeper slopes with high sediment loads.[30] Formations are common where water flow is slow, and there is a buildup of sediment in the river, causing changes in the direction of the river to create new, but often temporary, meandering channels.[31]

*Marte Vallis* is a valley in the *Amazonis Planitia* quadrangle of Mars, located at latitude 5.690° N and longitude 177.516° E. The valley is a proposed outflow channel, carved in the geological past by catastrophic release of water from aquifers beneath the Martian surface.[32] The valley displays noticeable braided flows analogous with those examined along the Tohachi Wash on the Moenkopi Plateau (Figure 10). Telemetry from TES at this particular flow reveals basalt and traces of sulfate, quartz, carbonates, feldspar, and hematite.[19] Conversely, the flow along the Tohachi Wash (latitude 35° 42' 41" N and longitude 111° 14' 07" W) similarly exhibits braided morphology, formed when water receding from the northeast slowed due to a buildup of sediment. Young terrace gravel, alluvium, and eolian deposits from Holocene and Pleistocene make up this braided river. Geologically, the river is dominated by clear quartz, basaltic ash fragments, siltstone mud curls, quartz sandstone grains and trace evidence of dark-red quartz, milky quartz, gray chert, biotite, gypsum and schist fragments.[3]

## ANALOG 7: OXBOWS

An oxbow is a crescent-shaped lake or river that forms when a vast meandering flow stops, creating a free-standing body of water or a U-shaped bend regardless of being cut off from the main waterway.[33] On the inside of the loop, the water travels more slowly, which leads to the deposition of silt. Meanwhile, water on the outside edges flows faster, which erodes the banks and widens the meander. Over time the loop of the meander expands until the neck disappears altogether.

Suggested meandering streambeds on *Aeolis Mensae* (latitude -5.58° S and longitude 153.551° E), for instance, exhibit a series of oxbow morphology analogous to those found on the Moenkopi Plateau (Figure 11). At *Aeolis Mensae*, some places on the stream show inverted relief, in which a stream bed could be a raised feature, rather than a valley. The inversion may be produced by the deposition of large rocks or by cementation that left the old channel as a raised ridge because the stream bed is more resilient to erosion.[34] Telemetry from TES at this particular oxbow reveals mostly traces of feldspar, sheet silicates, carbonates, and sulfate.[19] The oxbow on the Tohachi Wash (latitude 35° 51' 00" N and longitude 111° 11' 39" W), which is a member of the Glen Canyon Group (Lower to possibly Upper Triassic) is comprised of clear quartz, basaltic ash fragments, siltstone mud curls, quartz sandstone grains and trace evidence of dark-red quartz, gray chert, biotite, and gypsum.[3]

## ANALOG 8: CUT BANKS AND POINT BARS

In river morphology, as the water flows across the land, it erodes the soil and creates banks. Cut banks, in abundance along meandering streams, are located on the outside of a bend. They are cliff-shaped and molded by soil erosion as water flow collides with the river bank.[35] On the other hand, a point bar is a crescent-shaped depositional feature made of alluvium that accumulates on the inside of the bend. Point bars, like cut banks, are found in abundance in meandering streams and rivers.[36]

Cut banks and point bars are found along numerous channels on Mars, to include at *Hypanis Valles* (latitude 9.842° N and longitude 314.106° E). The *Hypanis Valles* are a set of channels in a 270-km valley in *Xanthe Terra* (Figure 12). The channels are burrowed between Middle Noachian highlands and feature volcanic, fluvial, undifferentiated impact, and basin materials. Some studies have proposed that long-lived flowing water carved these channels.[37]

Telemetry from TES at this particular Martian oxbow reveals mostly silicon, iron, thorium and traces of quartz, feldspar, hematite, sulfate, potassium and chlorine. The Little Colorado River on the Moenkopi Plateau exhibits cut bank and point bar morphology analogous to the channels on *Hypanis Valles*. In a particular point along the Little Colorado River (latitude 35° 47' 35" N and longitude 111° 19' 05" W), a series of erosion and deposition activity along the inside and outside of the bends reveal how cut banks and point bars are formed (Figure 12). The composition of samples along these bends are predominantly basaltic ash fragments, milky quartz, clear quartz, white quartz sandstone grains and traces of magnetite and siltstone mud curls.[3]

## CONCLUSION

An analysis of the HiRISE and CTX imagery of Mars has identified paleopotamologic features consistent with the visual record of the Moenkopi Plateau. Fluvial artifacts on Mars, such as tributaries, confluence, streamlined islands, terraces, alluvial fans, braided rivers, oxbows, cut banks, and point bars, were formed by the flow of water, as interpreted by the Earth analog. Moreover, accompanying fluvial interactions, such as drainage patterns, erosion, and deposition of sediments, are analogous on both planets. This investigation, therefore, has resolved the contention regarding paleopotamologic artifacts on Mars by demonstrating that the fluvial landscapes were formed by water. Accordingly, earlier patchwork investigations proposing glacial action and/or volcanic activity as the mechanism behind these fluvial artifacts on Mars can be ruled out.

## NOTES:

Though not conclusive or the primary purpose of our investigation, telemetry from TES and remote sensing data from the U.S. Geological Survey indicate that a particular mineral, quartz, has been detected on the alluvial features of Mars and the Moenkopi Plateau. Quartz is the second most copious mineral in the Earth's crust and forms in either igneous rocks or environments with geothermal waters. A recent study proposed that quartz found on Mars near *Antoniadi* Crater formed as a diagenetic product of hydrated amorphous silica, indicating there was once persistent water at *Antoniadi* Crater.[38]

## BIOGRAPHY

Antonio Paris, the Principal Investigator (PI) for this study, is the Chief Scientist at Planetary Sciences, Inc., a former Assistant Professor of Astronomy and Astrophysics at St. Petersburg College, FL, and a graduate of the NASA Mars Education Program at the Mars Space Flight Center, Arizona State University. He is the author of "*Mars: Your Personal 3D Journey to the Red Planet*". His latest peer-reviewed publication includes *Prospective Lava Tubes at Hellas Planitia* – an investigation into leveraging lava tubes on Mars to provide crewed missions protection from cosmic radiation. Prof. Paris is a professional member of the Washington Academy of Sciences and the American Astronomical Society.

## FIELD RESEARCH CONTRIBUTOR

Laurence A. Tognetti recently graduated from Arizona State University with a Master's Degree in Geological Sciences with thesis research focusing on geomorphological processes of the Martian surface. As a Field Researcher for the Planetary Sciences Inc., he operated the UAV in situ on the Moenkopi Plateau and assisted in imagery analysis.

Approved and Accepted by the Washington Academy of Sciences on 30 April 2020

# APPENDIX 1

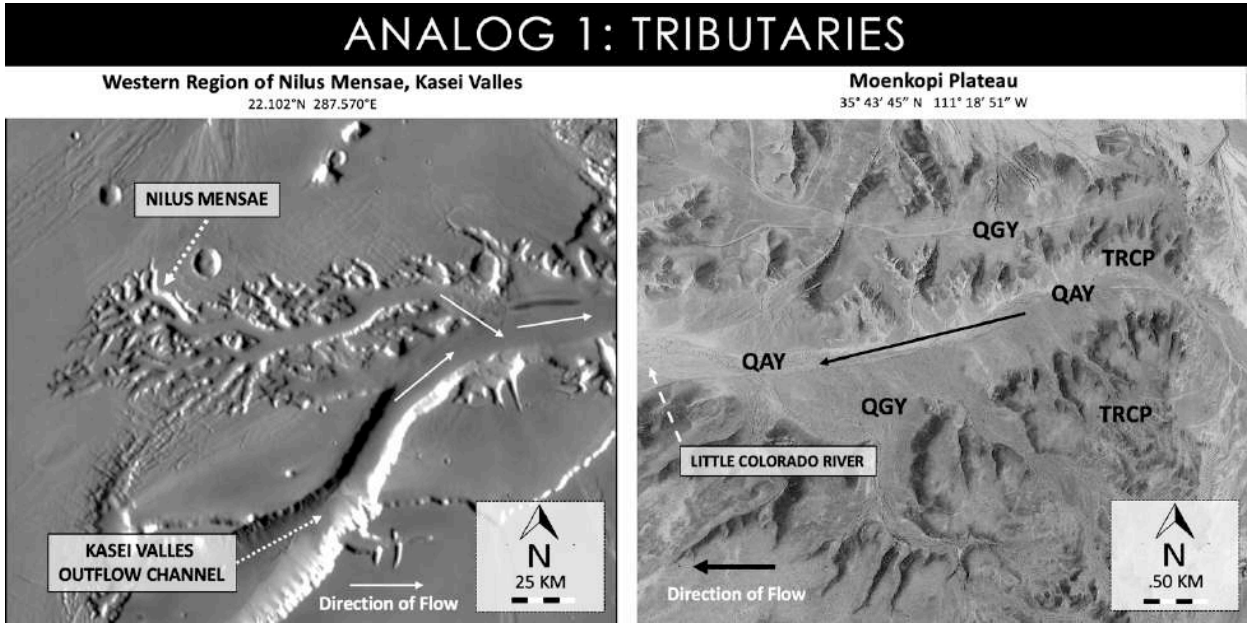

Figure 5: Nilus Mensae (Credit: NASA) and Moenkopi Plateau (Credit: Planetary Sciences, Inc. UAV)

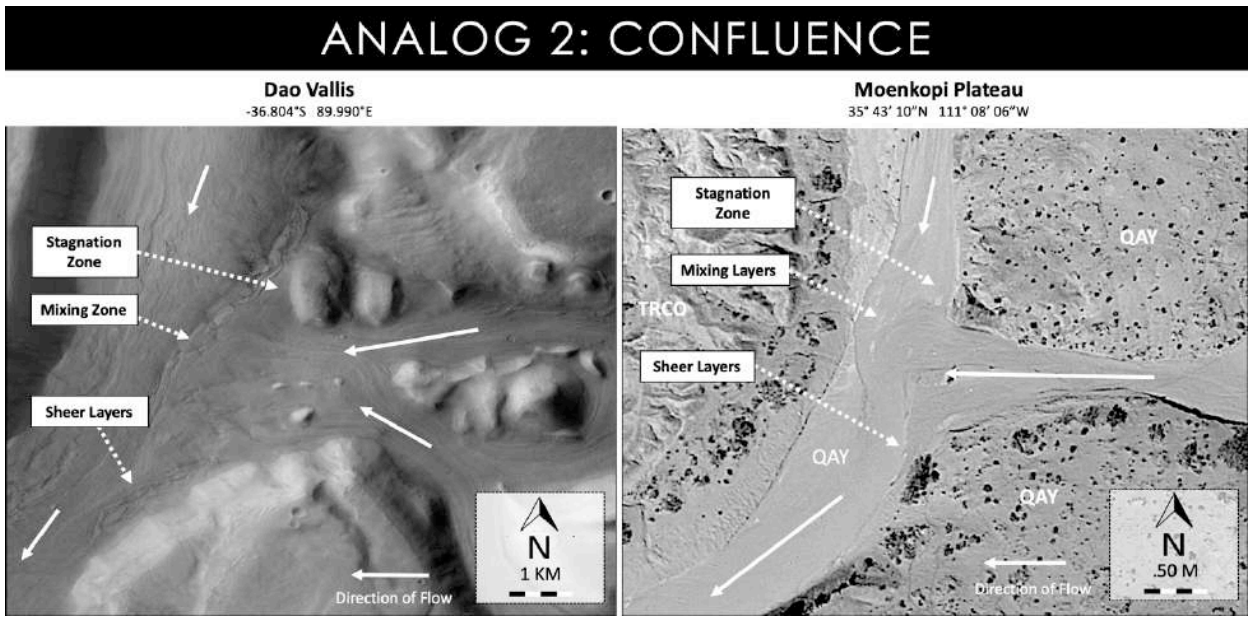

Figure 6: Dao Vallis (Credit: NASA) and Moenkopi Plateau (Credit: USGS)

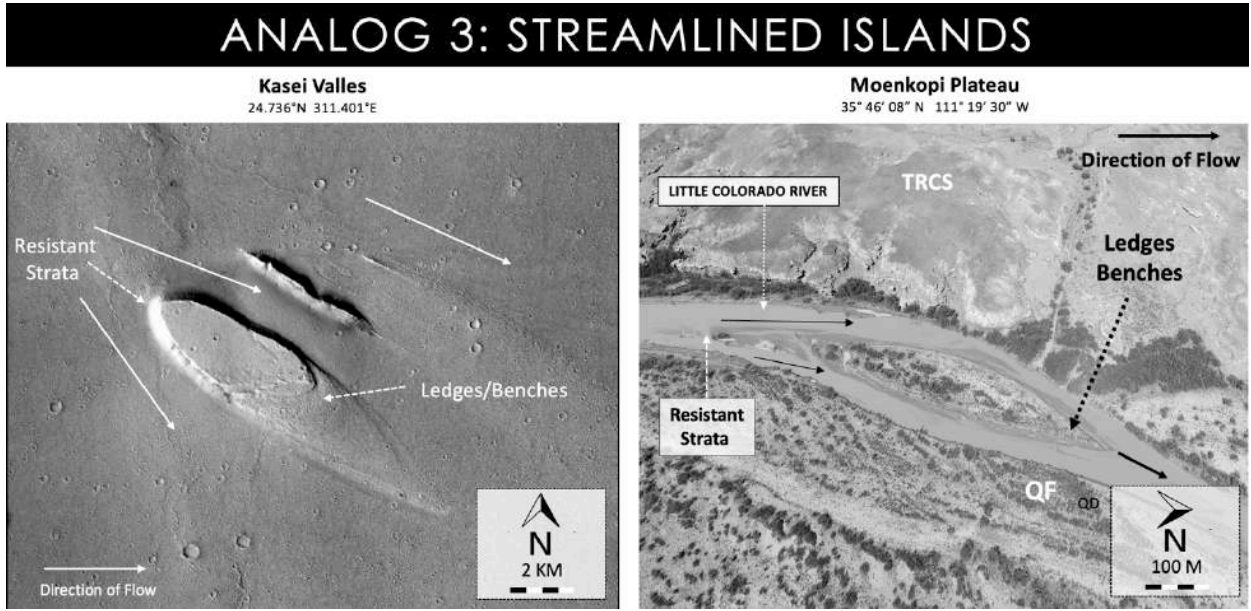

Figure 7: Kasei Valles (Credit: NASA) and Moenkopi Plateau (Credit: Planetary Sciences, Inc. UAV)

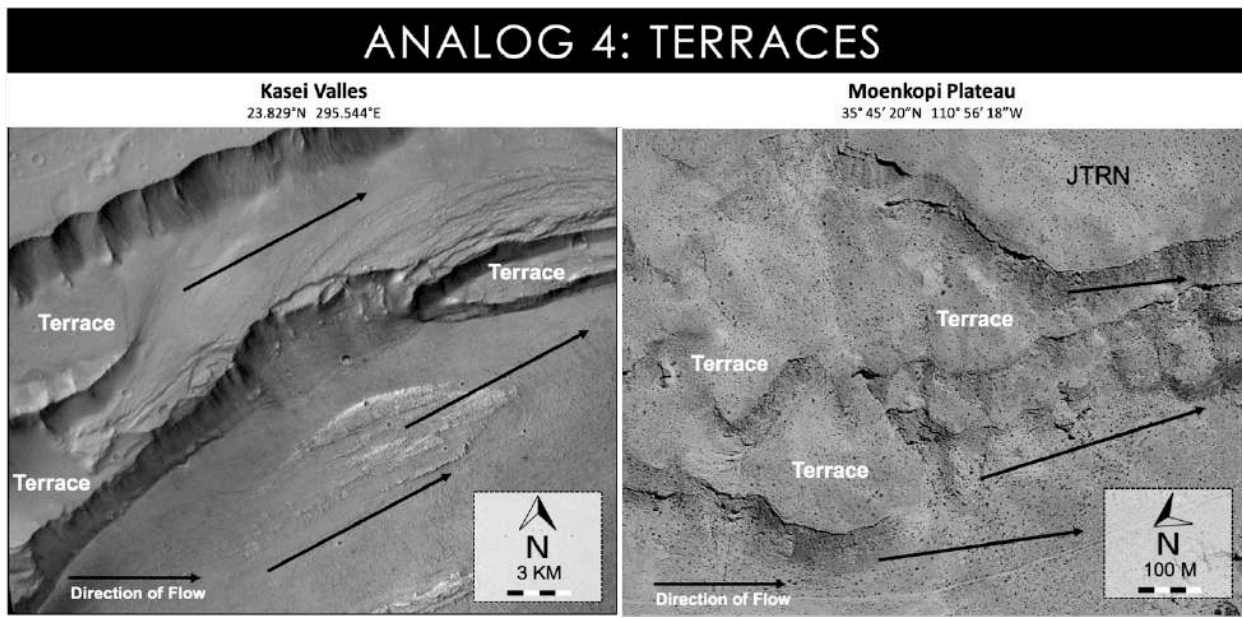

Figure 8: Kasei Valles (Credit: NASA) and Moenkopi Plateau (Credit: USGS)

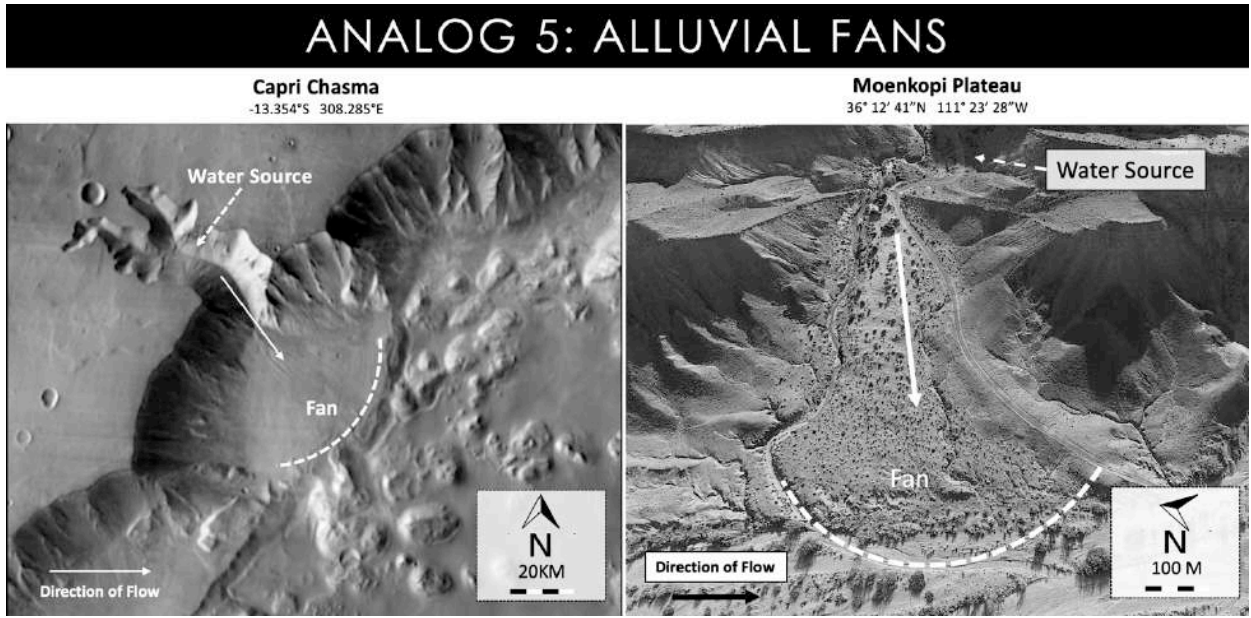

Figure 9: Capri Chasma (Credit: NASA) and Moenkopi Plateau (Credit: Planetary Sciences, Inc. UAV)

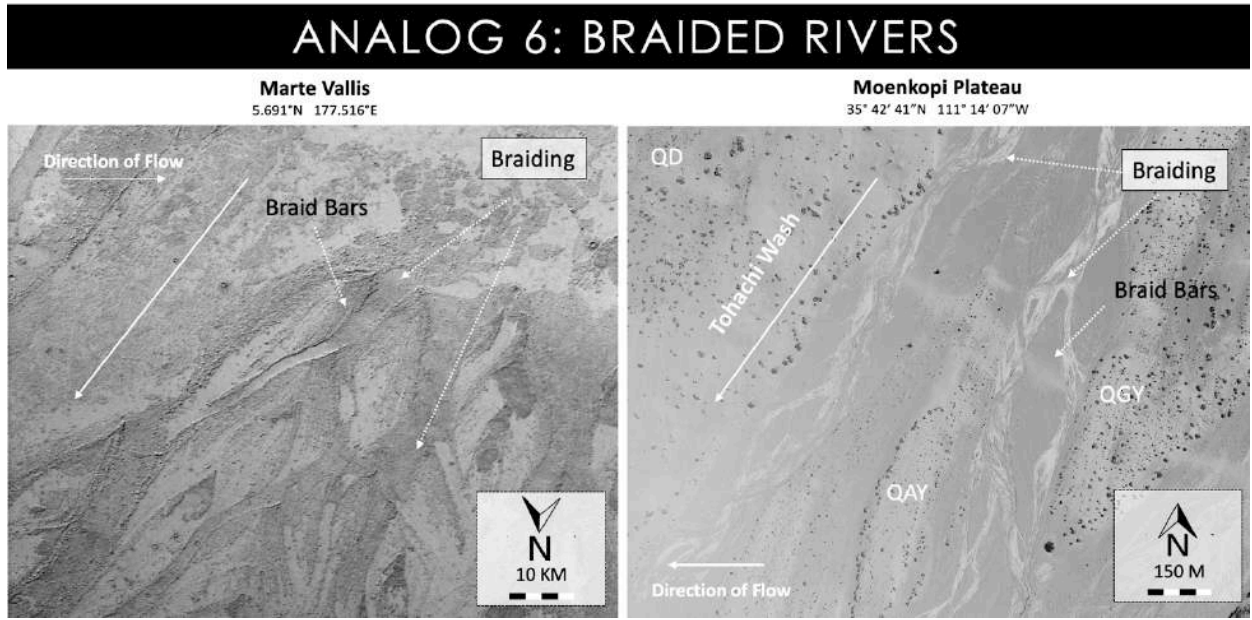

Figure 10: Marte Vallis (Credit: NASA) and Moenkopi Plateau (Credit: Planetary Sciences, Inc. UAV)

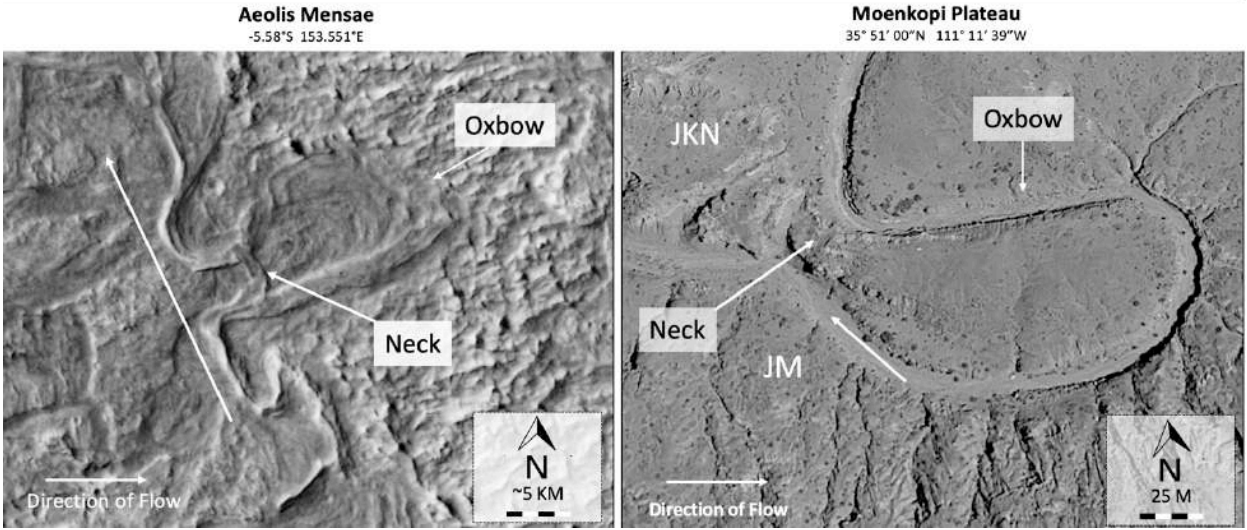

Figure 11: Aeolis Mensae (Credit: NASA) and Moenkopi Plateau (Credit: USGS)

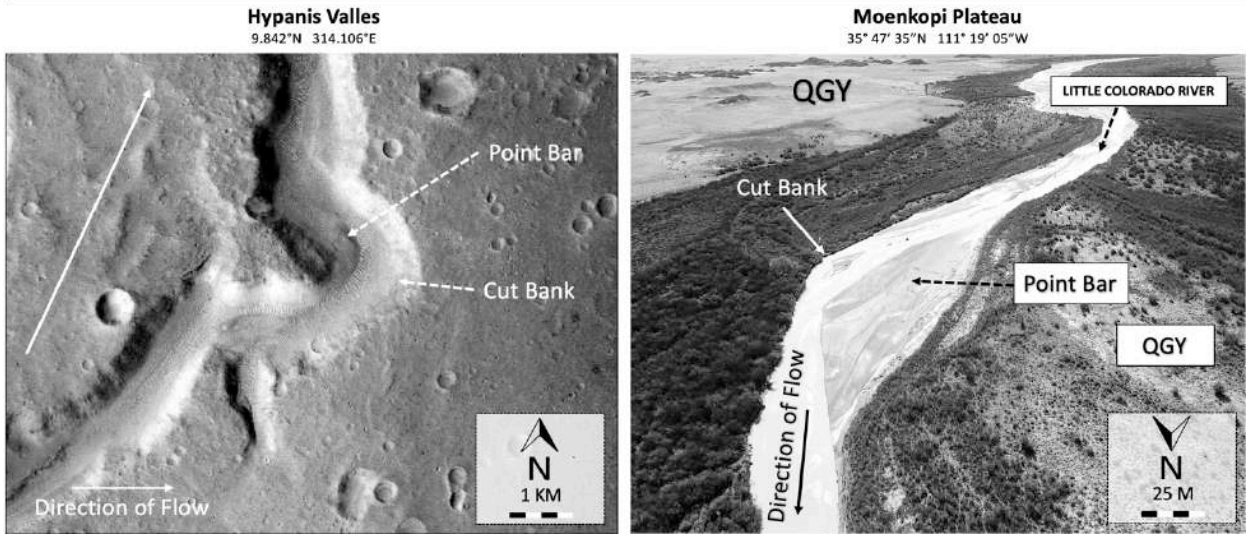

Figure 12: Hypanis Valles (Credit: NASA) and Moenkopi Plateau (Credit: Planetary Sciences, Inc. UAV)